\documentclass[preprint,onecolumn,nofootinbib]{revtex4}
\pdfoutput=1
\usepackage[colorlinks=true,linkcolor=blue,urlcolor=blue,filecolor=black,citecolor=red,pdfstartview=FitV,pdftitle={},pdfsubject={},pdfkeywords={},pdfpagemode=None,bookmarksopen=true]{hyperref}
\usepackage{graphicx}
\usepackage{amsmath}
\usepackage{amsfonts}
\usepackage{amssymb,ulem}
\usepackage{color}%
\usepackage{tikz}
\usepackage{dcolumn}
\usepackage{xcolor}
\usepackage{subcaption}
\usepackage{booktabs}
\usepackage[capitalize]{cleveref} 
\begin{document}
\title{
	Diagnosing Critical Behavior in AdS Einstein-Maxwell-Scalar Theory via Holographic Entanglement Measures
}
\author{Zhe Yang $^{1,2}$}
\email{yzar55@stu2021.jnu.edu.cn}
\author{GuangZai Ye $^{1}$}
\email{photony@stu2022.jnu.edu.cn}
\author{Jian-Pin Wu $^{2}$}
\email{jianpinwu@yzu.edu.cn}
\thanks{Corresponding author}
\author{Peng Liu $^{1}$}
\email{phylp@email.jnu.edu.cn}
\thanks{Corresponding author}

\affiliation{
   $^1$ Department of Physics and Siyuan Laboratory, Jinan University, Guangzhou 510632, China\\
  $^2$ Center for Gravitation and Cosmology, College of Physical Science and Technology, Yangzhou University, Yangzhou 225009, China\\
  }

\begin{abstract}

	We investigate the holographic mixed-state entanglement measures in the Einstein-Maxwell-Scalar (EMS) theory. Several quantities are computed, including the holographic entanglement entropy (HEE), mutual information (MI), entanglement wedge cross-section (EWCS), and butterfly velocity ($v_B$). Our findings demonstrate that these measures can effectively diagnose phase transitions. Notably, EWCS and MI, as mixed-state entanglement measures, exhibit behavior opposite to that of the HEE. Additionally, we study the butterfly velocity, a dynamic quantum information measure, and observe that it behaves differently from the static quantum information measures. We analyze the butterfly velocity and find that its non-monotonic behavior arises from the competition between two contributions in its expression, which the analytic structure suggests may be correlated with distinct physical interpretations. Moreover, we examine the scaling behavior of the holographic entanglement measures and find that all the critical exponents are equal to $1$, which is twice that of the scalar field. We also explore the inequality between EWCS and MI, noting that the growth rate of MI consistently exceeds that of EWCS during phase transitions. These features are expected to be universal across thermodynamic phase transitions, with the inequalities becoming more significant as one moves away from the critical point.

\end{abstract}
\maketitle
\tableofcontents

\section{Introduction}

Quantum entanglement stands as one of the most fundamental phenomena in quantum mechanics, exhibiting properties fundamentally distinct from classical physics. Over recent years, it has become an essential tool for exploring quantum information, condensed matter physics, and quantum gravity \cite{Osterloh2002,RevModPhys.80.517,Ryu:2006bv,Lewkowycz2013,Veronika2007}. Notably, quantum information measures serve as probes for phase transitions and play a pivotal role in the emergence of spacetime \cite{TatsumaNishioka2007,KLEBANOV2008274,AriPakman_2008,Zeng2016,Mahapatra2019}.

Among various quantum information measures, entanglement entropy (EE) remains the most widely used for quantifying entanglement. However, EE fails to capture mixed-state entanglement---a more common scenario than pure states. To address this limitation, alternative measures such as reflected entropy, mutual information (MI), and entanglement of purification (EOP) have been developed \cite{Vidal:2002zz,Horodecki:2009zz,plenio2005logarithmic}. These measures have proven effective in diagnosing various phase transitions, including quantum, topological, and superconducting transitions \cite{Osterloh2002,RevModPhys.80.517,Wen:2006topo}, revealing a deep connection between entanglement and critical phenomena. In condensed matter physics, many important phase transitions occur in strongly correlated systems---high-temperature superconductors, heavy fermions, and the fractional quantum Hall effect \cite{Si:2010xrl, neto2001charge, gegenwart2008quantum, bolotin2009observation}. Studying entanglement in these systems remains exceptionally challenging, necessitating novel theoretical approaches.

The gauge/gravity duality has emerged over the past two decades as a powerful framework for studying strongly correlated systems \cite{SusskindHolo,MaldacenaLarge}. Within this framework, a strongly correlated quantum system maps to a classical gravitational theory, enabling the construction of holographic duals for various entanglement measures. The most prominent example is holographic entanglement entropy (HEE), corresponding to the minimal surface area in the bulk \cite{Ryu:2006bv}. HEE has been extensively applied to probe thermodynamic and quantum phase transitions \cite{Caceres:2015vsa,Liu:2020blk, Cai:2012nm, Ling:2015dma, Ling:2016wyr,Liu2019,LIU2019155,Fu_2021,Gong_2021,Huang:2019zph}. More recently, the entanglement wedge cross-section (EWCS) was introduced as a novel measure \cite{Takayanagi:2017knl, Umemoto:2018jpc}, serving as the holographic dual of reflected entropy, logarithmic negativity, and balanced partial entanglement \cite{Kudler-Flam:2018qjo, Dutta:2019gen, Jokela:2019ebz,Ling:2021vxe}. Studies have demonstrated that EWCS provides a robust measure of mixed-state entanglement \cite{Huang:2019zph, Chen:2021bjt, Yang:2023wuw, Liu:2020blk, Liu:2021rks,Chen:2024wpt,Li:2023edb,Liu:2023rhd, Chowdhury:2021idy,Paul:2024lmd,Karan:2023hfk,Jain:2022hxl,ChowdhuryRoy:2022dgo,Jain:2020rbb}. Beyond these static measures, the butterfly velocity $v_B$ captures the dynamical spread of quantum information and characterizes chaos in the system \cite{PhysRevLett.117.091602}. This dynamical probe has been studied across various holographic models \cite{Shenker:2013pqa, Blake:2016wvh,Liu:2021stu,Ling:2016ibq,Ling:2016wuy,Lilani:2025wnd,Zhao:2025gej}, complementing the static entanglement measures in characterizing phase transitions.

The Einstein-Maxwell-Scalar (EMS) theory represents a paradigmatic holographic model exhibiting spontaneous scalarization as temperature or coupling constant varies. Upon crossing the critical point, the system undergoes a phase transition from the normal to the scalarized state. While extensive studies have examined HEE in this context, a comprehensive analysis of EWCS as a mixed-state entanglement probe remains absent. Our previous work on holographic p-wave superconductors revealed an inequality between the growth rates of EWCS and MI \cite{Yang:2023wuw}, raising the question of whether this inequality persists in the EMS theory. Furthermore, butterfly velocity---as a dynamical quantum information measure---exhibits distinct behavior from static measures like HEE and EWCS. Since HEE receives significant contributions from thermal entropy in mixed-state systems while EWCS more faithfully captures quantum correlations, understanding the behavior of $v_B$ in this context fills a critical gap. This paper presents a systematic investigation of holographic entanglement measures in EMS theory.

This paper is organized as follows. In \cref{sec:2}, we introduce the EMS model and the relevant holographic quantum information measures: HEE, MI, EWCS, and butterfly velocity. \cref{sec:3} presents numerical results and analyzes the scaling behavior of these measures. In \cref{sec:ineq}, we examine the inequality between the growth rates of MI and EWCS. We conclude with a summary and outlook in \cref{sec:4}.

\section{Holographic setup for Einstein-Maxwell-Scalar theory and holographic information-related quantities}
\label{sec:2}

\subsection{The AdS Einstein-Maxwell-Scalar Model}
The action of the Einstein-Maxwell-scalar theory in the AdS spacetime is \cite{Xiong:2022ozw,Zhang:2021etr},
\begin{equation}
	S=\int \mathrm{d}^{4}x\sqrt{-g}\left[R +\frac{6}{L^{2}} - \nabla _{\mu }\phi\nabla ^{\mu }\phi -\frac{1}{2}f(\phi )F_{\mu \nu }F^{\mu \nu } \right],
\end{equation}
where $R$ is the Ricci scalar, $L$ is the AdS length scale, and $\phi $ is a real scalar field. $A_\mu$ is the gauge field and the field strength $F_{\mu \nu }=\nabla _{\mu }A_{\nu }-\nabla _{\nu }A_{\mu }$. $f(\phi )$ is the coupling function between electromagnetic and scalar fields. In this paper, we define it as $f(\phi)=e^{-b\phi ^{2}}$ and $b$ is a dimensionless coupling constant. The equation of motions (EOMs) of this action are,
\begin{align}
	 & R_{\mu \nu }-\frac{1}{2}Rg_{\mu \nu }-\frac{3}{L^{2}}g_{\mu \nu }=T^{\phi }_{\mu \nu }+f(\phi )T^{A}_{\mu \nu }\label{eq:space time}, \\
	 & \nabla_{\mu }\nabla ^{\mu }\phi =\frac{1}{4}\frac{\mathrm{d}f(\phi )}{\mathrm{d}\phi }F_{\mu \nu }F^{\mu \nu }\label{eq:real field},  \\
	 & \nabla_{\mu }(f(\phi)F^{\mu \nu })=0, \label{eq:gauge field}
\end{align}
where the energy-momentum tensor
\begin{align}
	T^{\phi }_{\mu \nu } & =\nabla _{\mu }\phi \nabla _{\nu }\phi -\frac{1}{2}g_{\mu \nu }\nabla _{\rho }\phi \nabla ^{\rho }\phi, \label{eq:emTp}\\
	T^{A}_{\mu \nu }     & =F_{\mu \rho }{F^{\rho }}_{\nu }-\frac{1}{4}g_{\mu \nu }F_{\rho \sigma }F^{\rho \sigma }. \label{eq:emTA}
\end{align}
We solve the EOM with this ansatz,
\begin{align}
	\mathrm{d}s^{2}= & \frac{1}{z^{2}}\left (-(1-z)p(z)U(z)\mathrm{d}t^{2}+\frac{\mathrm{d}z^{2}}{(1-z)p(z)U(z)}+V(z)(\mathrm{d}x^{2}+\mathrm{d}y^{2})\right ), \\
	A_{\mu}=           & \mu(1-z)a(z)\mathrm{d}t.
  \label{eq:ansatz}
\end{align}
where $\mu$ is the chemical potential of the gauge field and $p(z)= 1+z+z^{2}-\mu^{2}z^{3}/2$. We set $ z = r_h / r $, where the radial coordinate \( z \) ranges from 0 to 1, with $ z = 0 $ representing the AdS boundary and $ z = 1 $ representing the horizon.
The quantities $U, V, a,\phi$ are the functions of $z$, which can be obtained by solving the EOM. Additionally, the ansatz will return to AdS-RN black brane when we set $a=U=V=1$, and $\phi=0$.
\begin{figure}
	\centering
	\includegraphics[width=0.5\textwidth]{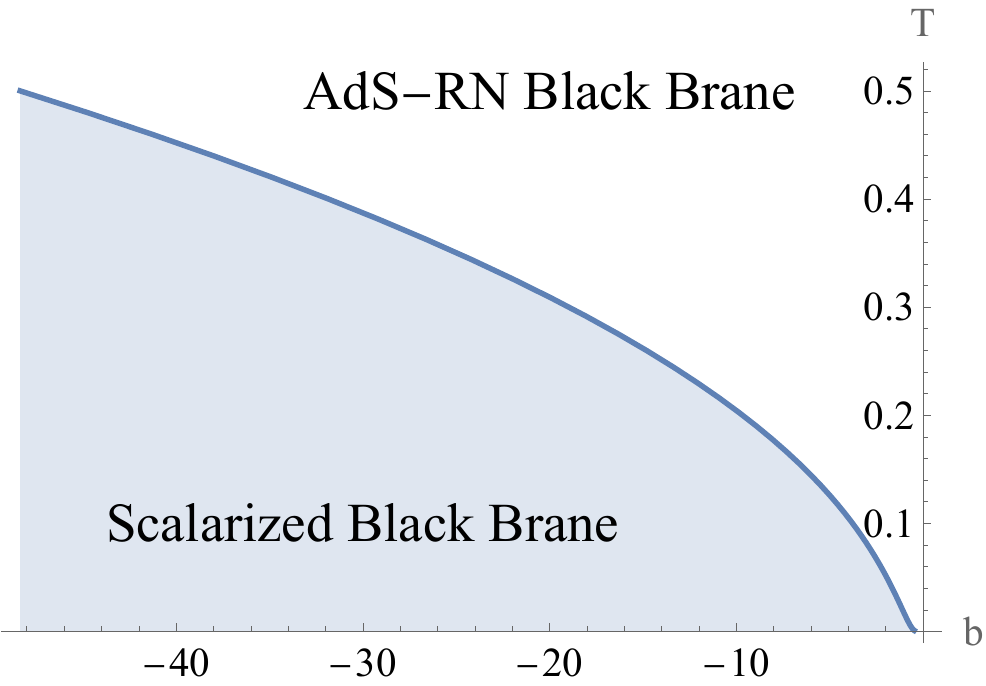}
	\caption{Phase diagram of thermal phase transition in the EMS model.}
	\label{fig:phase}
\end{figure}

The Hawking temperature of this system is given by $\tilde{T}=\frac{6-\mu^{2}}{8\pi}$. Notably, the system remains invariant under the following rescaling:
\begin{equation}
	\begin{aligned}
		(t,x,y)\to & \alpha^{-1} (t,x,y), & \qquad & V\to \alpha^{2}V,              \\
		\mu\to     & \alpha \mu,          & \qquad & \tilde{T}\to \alpha \tilde{T}.
	\end{aligned}
\end{equation}
In this paper, we set $\mu$ as the scaling unit, making the dimensionless Hawking temperature $T=\tilde{T}/\mu$. It is evident that by varying the coupling constant $b$ and the Hawking temperature $T$, the system transitions from an AdS-RN black brane to a scalarized black hole. The phase transition of this system is illustrated in Fig. \ref{fig:phase}.

\subsection{Holographic Information-Related Quantities}

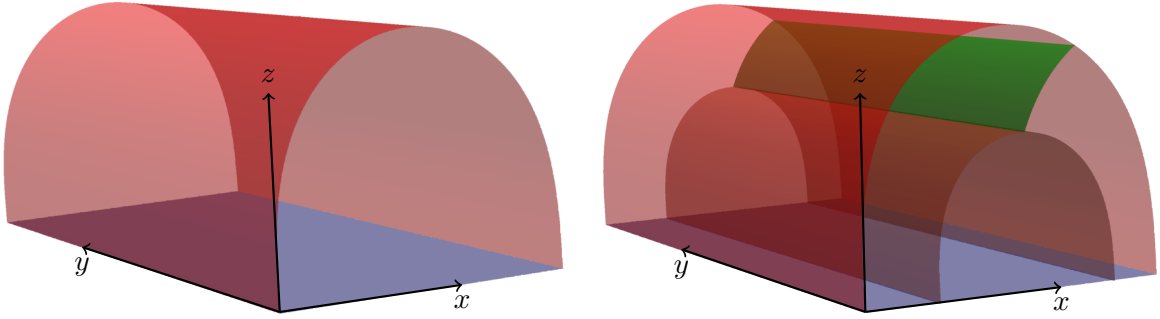
\begin{figure}
	\begin{tikzpicture}[scale=1]
		\node [above right] at (0,0) {\includegraphics[width=7.5cm]{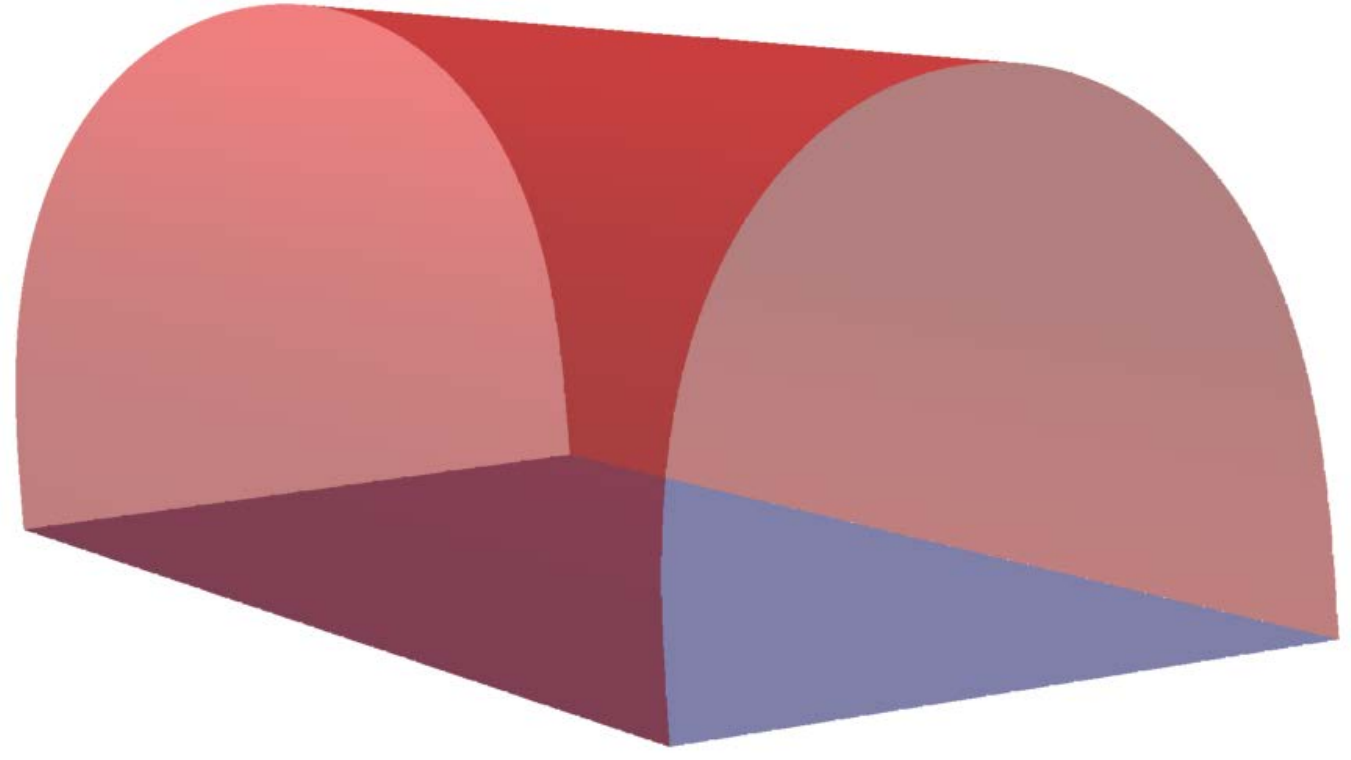}};
		\draw [right,->,thick] (3.85, 0.22) -- (6.25, 0.58) node[below] {$x$};
		\draw [right,->,thick] (3.85, 0.22) -- (1.25, 1.08) node[below] {$y$};
		\draw [right,->,thick] (3.85, 0.22) -- (3.7, 3.125) node[above] {$z$};
	\end{tikzpicture}
	\begin{tikzpicture}[scale=1]
		\node [above right] at (0,0) {\includegraphics[width=7.5cm]{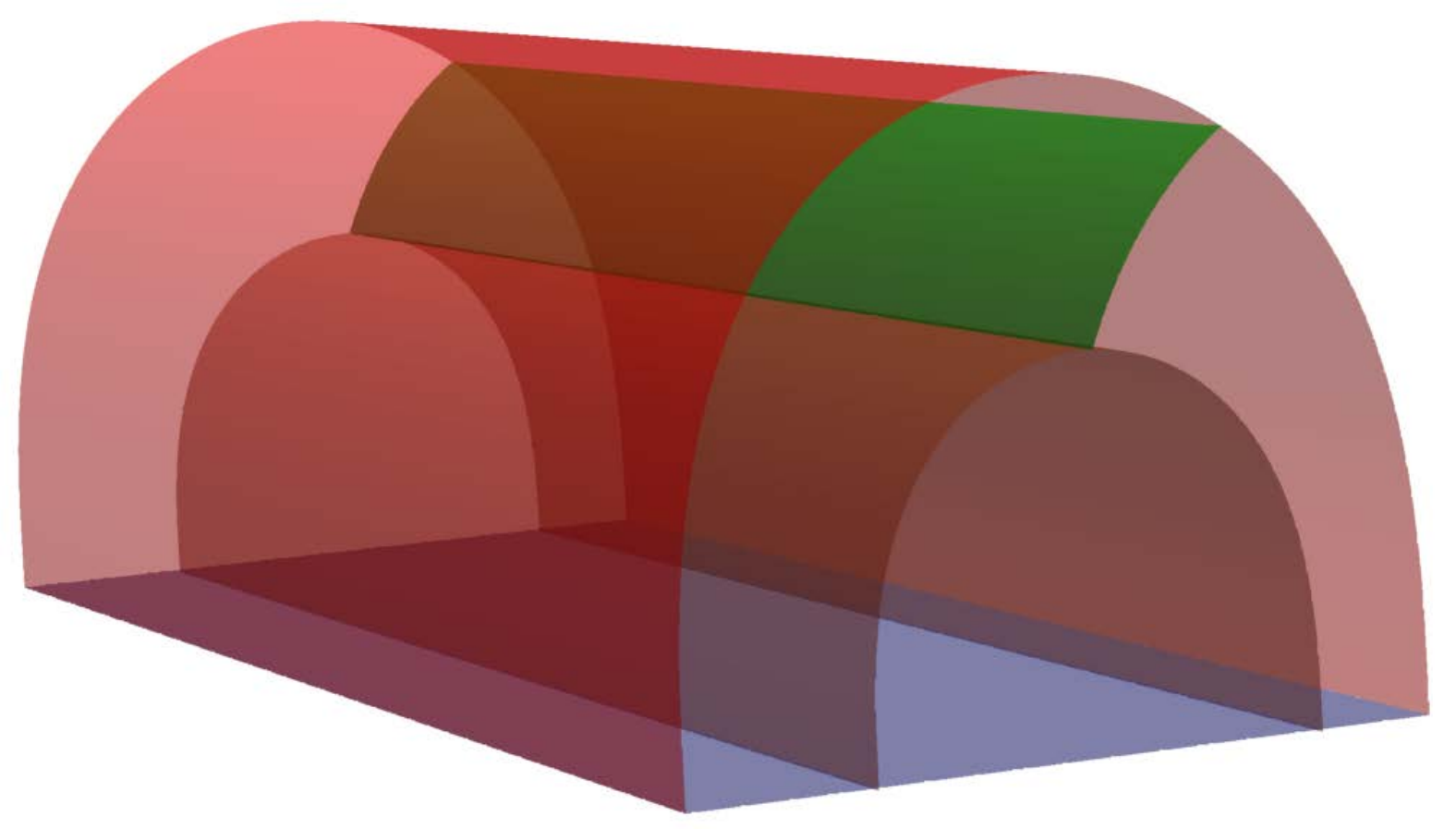}};
		\draw [right,->,thick] (3.67, 0.22) -- (6.25, 0.55) node[below] {$x$};
		\draw [right,->,thick] (3.67, 0.22) -- (1.25, 1.05) node[below] {$y$};
		\draw [right,->,thick] (3.67, 0.22) -- (3.6, 3.125) node[above] {$z$};
	\end{tikzpicture}
	\caption{The left plot: The minimum surface for a given width $w$.
		The right plot: The minimum cross-section (green surface) of the entanglement wedge.}
	\label{fig:msd1}
\end{figure}

Quantum entanglement fundamentally distinguishes quantum from classical physics. Among the various measures of entanglement, entanglement entropy (EE) stands out as the most widely adopted. For a density matrix $\rho_A$, EE is defined as \cite{Eisert:2008ur}
\begin{equation}
	\rho_{A}= \text{Tr}_{B}(\rho),\qquad S_{A}=-\text{Tr}(\rho_{A}\log\rho_{A}).
\end{equation}
In the holographic framework, the EE of a boundary region $A$ equals the area of the corresponding minimal surface in the bulk---the holographic entanglement entropy (HEE), illustrated in the left plot of Fig.~\ref{fig:msd1}. Specifically,
\begin{equation}\label{eq:hee}
	S_{A}=\frac{\gamma_{A}}{4G_{N}}.
\end{equation}
where $G_{N}$ is the Newton constant and $\gamma_{A}$ is the area of the minimal surface that corresponds to the region $A$ on the boundary \cite{Ryu:2006bv}. However, the asymptotic behavior of the AdS boundary causes the HEE to diverge, which can be renormalized by subtracting the divergent term. Therefore, the HEE $S_E$ in this paper should be regarded as a renormalized holographic entanglement entropy and can take negative values.

We consider an infinite strip along the $y$-direction in a homogeneous background geometry:
\begin{equation}
	\mathrm{d}s^{2}=g_{tt}\mathrm{d}t ^{2}+g_{zz}\mathrm{d}z^{2}+g_{xx}\mathrm{d}x^{2}+g_{yy}\mathrm{d}y^{2}.
\end{equation}
As illustrated in \cref{fig:msd1}, the minimal surface area depends on the strip width $w$:
\begin{equation}\label{eq:area}
	A_{\gamma_A}=L_y\int^{w}_0\sqrt{g_{yy}(g_{xx}+g_{zz}z'(x)^2)}dx,
\end{equation}
where $L_y=\int dy$ and the strip width is $w=\int dx$. For an infinite strip along the $y$-direction, $L_y$ factors out. Minimizing Eq.~\eqref{eq:area} via the Euler-Lagrange equation yields
\begin{equation}
	\label{eq:eomA}
	2g_{yy}g_{zz} z'(x)^2 g'_{xx}+g_{xx}(g_{yy}(-2g_{zz}z''(x)-z'(x)^2g'_{zz}+g'_{xx})+g_{zz}z'(x)^2 g'_{yy})+g_{xx}^2g'_{yy}=0.
\end{equation}
This equation enables numerical computation of the HEE \cite{Liu2019}. However, in mixed-state systems, HEE receives substantial contributions from thermal entropy \cite{Ling:2015dma}, motivating the search for alternative entanglement measures.

\begin{figure}
	\centering
        \includegraphics[width = 0.5\textwidth]{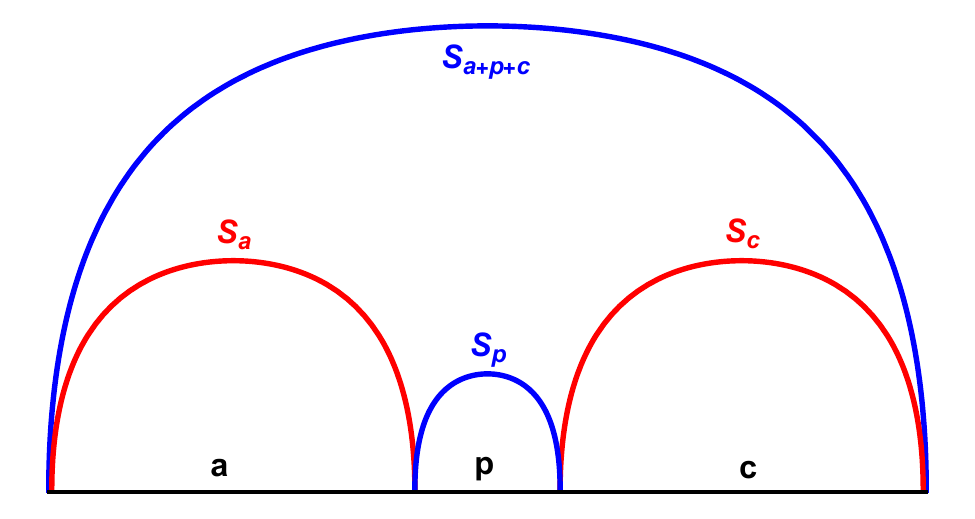}
	\caption{
		The Illustration of holographic mutual information.
	}
	\label{fig:MI}
\end{figure}
 Mutual information (MI) provides a more suitable measure for mixed-state entanglement \cite{Horodecki:2009zz,Vidal:2002zz}. Unlike EE, which remains nonzero even for unentangled product states $\mathcal{H_A}\otimes \mathcal{H_B}$, MI vanishes for such states by construction. We consider a bipartite system with subsystems $a$ and $c$ separated by region $p$ (Fig.~\ref{fig:MI}). The holographic MI reads \cite{chuang:2002}
\begin{equation}
	I(a,c)= S(a)+S(c)-\text{min}(S(a \cup c)).
 \label{eq:mi}
\end{equation}
where $S(a)$ and $S(c)$ denote the HEE of subsystems $a$ and $c$, respectively. A positive MI indicates the presence of entanglement. Since MI is non-negative by definition, the system undergoes a disentangled phase transition when MI decreases to zero. Nevertheless, MI can still be influenced by thermal entropy in certain configurations \cite{Huang:2019zph}, prompting interest in additional mixed-state measures.

The EWCS has emerged as a promising mixed-state entanglement measure \cite{Takayanagi:2017knl, Umemoto:2018jpc}, defined as the minimal cross-sectional area within the entanglement wedge (right plot of Fig.~\ref{fig:msd1}). It serves as the holographic dual of various quantum information measures \cite{Kudler-Flam:2018qjo, Dutta:2019gen, Jokela:2019ebz}:
\begin{equation}\label{eq:eqew}
	E_{W}(\rho _{ab})=\min_{\Sigma_{ab}}\left(\frac{\text{Area}(\Sigma_{ab})}{4G_{N}}\right).
\end{equation}

\begin{figure}
	\centering
        \includegraphics[width = 0.6\textwidth]{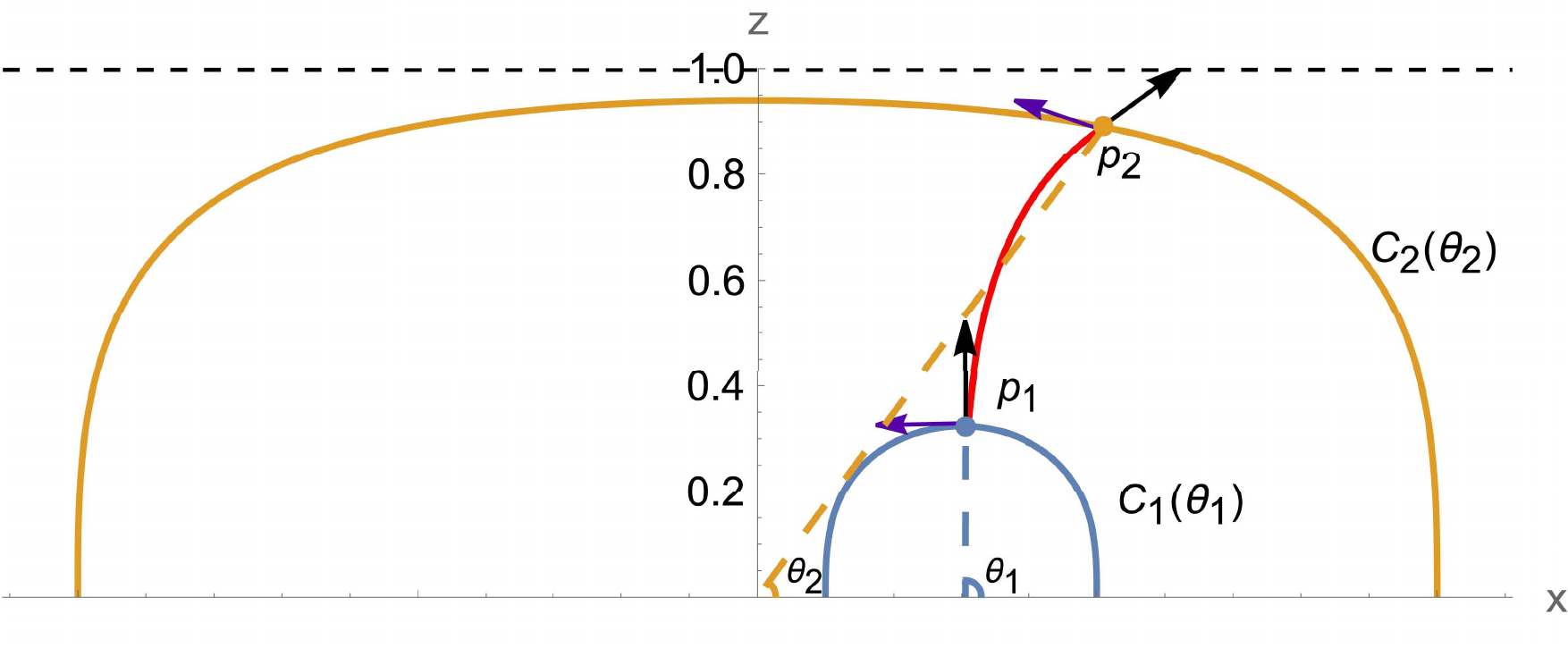}
	\caption{
		The illustration depicting the calculation of EWCS, and the red surface representing the minimum cross-section.
	}
	\label{fig:cartoon4eop}
\end{figure}
However, the calculation of EWCS remains challenging. First, the EOM for the minimum surface in the EWCS are highly nonlinear, making them difficult to solve. Second, finding the minimum surface is essentially a quadratic minimization problem, as the EWCS represents a global minimum cross-section located on the minimum surface within the entanglement wedge. This makes the search for the global minimum cross-section particularly challenging. Third, the coordinates near the AdS boundary ($z \to 0$) are singular, requiring high numerical precision to ensure accuracy.

Next, we will briefly introduce the algorithm for calculating the EWCS, particularly for the asymmetric EWCS, which involves more diverse configurations \cite{Liu2019}. In this paper, we consider a bipartite system $a \cup c$ divided by region $p$. It should be noted that EWCS exists only when the MI is greater than zero. As illustrated in Fig. \ref{fig:cartoon4eop}, the minimum surfaces of regions $c$ and $a + p + c$ can be represented as $(C_1(\theta_1), C_2(\theta_2))$, respectively. The cross-section intersects these minimum surfaces at points $p_1$ and $p_2$, and the area of the cross-section can be calculated as follows:
 \begin{equation}
 A=\int_{C_{p_1,p_2}}\sqrt{g_{xx}g_{yy}x'(z)^2+g_{zz}g_{yy}}dz.
 \label{eq:ewcs}
 \end{equation}
Varying Eq.~\eqref{eq:ewcs}, we obtain the EOM of the cross-section:
\begin{equation}
x'(z)^3\left( \frac{g_{xx}g'_{yy}}{2g_{yy}g_{zz}}+\frac{g'_{xx}}{2g_{zz}}\right)+x'(z)\left( \frac{g'_{xx}}{g_{xx}}+\frac{g'_{yy}}{2g_{yy}}-\frac{g'_{zz}}{2g_{zz}}\right)+x''(z)=0.
\end{equation}
It needs to be noted that the global minimum cross-section is orthogonal to the entanglement wedge, which implies that
\begin{equation}
  \left \langle \frac{\partial}{\partial_z},\frac{\partial}{\partial \theta_1}\right \rangle_{p_1}=0, \quad
  \left \langle \frac{\partial}{\partial_z},\frac{\partial}{\partial \theta_2}\right \rangle_{p_2}=0
\end{equation}
where $\langle \cdot,\cdot\rangle$ represents the vector product with metric $g_{\mu\nu}$. We can normalize the orthogonal relation,
\begin{equation}
  \label{eq:bc1}
  Q_1(\theta_1,\theta_2) \equiv\left.\frac{\langle \frac{\partial}{\partial z},\frac{\partial}{\partial \theta_1}\rangle}{\sqrt{\langle \frac{\partial}{\partial z},\frac{\partial}{\partial z}\rangle \langle\frac{\partial}{\partial \theta_1},\frac{\partial}{\partial \theta_1}\rangle}} \right |_{p_1}=0,\quad
  Q_2(\theta_1,\theta_2) \equiv\left.\frac{\langle \frac{\partial}{\partial z},\frac{\partial}{\partial \theta_2}\rangle}{\sqrt{\langle \frac{\partial}{\partial z},\frac{\partial}{\partial z}\rangle \langle\frac{\partial}{\partial \theta_2},\frac{\partial}{\partial \theta_2}\rangle}} \right |_{p_2} =0.
\end{equation}
The cross-section endpoints lie on the two minimal surfaces $(C_1(\theta_1), C_2(\theta_2))$ at parameters $(\theta_1, \theta_2)$. Imposing the boundary conditions Eq.~\eqref{eq:bc1} locates the global minimum; we employ the Newton-Raphson method to find $(\theta_1, \theta_2)$ and thereby compute the asymmetric EWCS.

Unlike the static measures discussed above, the butterfly velocity $v_B$ captures the dynamical spread of quantum information. It quantifies the speed of chaos propagation in many-body systems and, in the holographic picture, is determined by localized shockwaves on the black hole horizon \cite{Shenker:2013pqa, Blake:2016wvh}. This dynamical quantity encodes the spreading and causal structure of the system, revealing deep connections between entanglement and geometry. For general anisotropic black branes \cite{Ling:2016ibq},
\begin{equation}
	v_{B}=\sqrt{\frac{-2 \pi  T \mu  V_y(z)}{V_{y}(z)(V_{x}'(z)-2V_{x}(z))+V_{x}(z)(V'_{y}(z)-2V_{y}(z))}}\Bigg| _{z=1}.
\end{equation}
For an isotropic EMS model, this simplifies to
\begin{equation}\label{eqt:vb}
	v_B=\sqrt{\frac{\pi T \mu }{2V(z)-V'(z)}}\Bigg| _{z=1}.
\end{equation}
Thus $v_B$ depends on the Hawking temperature $T$, chemical potential $\mu$, and background geometry $V(z)$, enabling a direct comparison with static entanglement measures across the phase transition.

\section{The computation of holographic entanglement measures}\label{sec:3}

With the holographic setup established in the previous section, we now present the numerical results for various entanglement measures and analyze their behavior during the phase transition.

\subsection{The holographic entanglement entropy and the mutual information}
\begin{figure}
\centering
\includegraphics[width=0.45\textwidth]{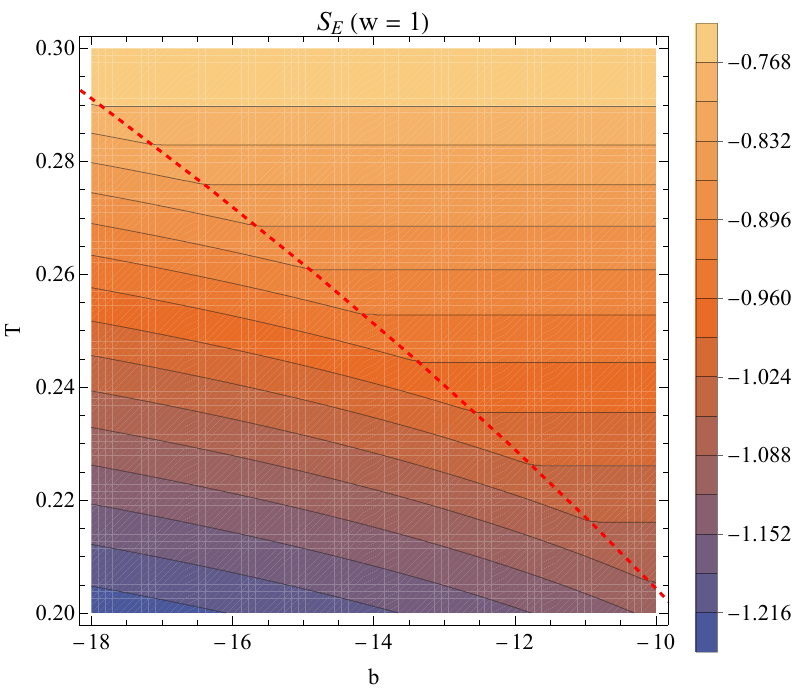}\quad
\includegraphics[width=0.45\textwidth]{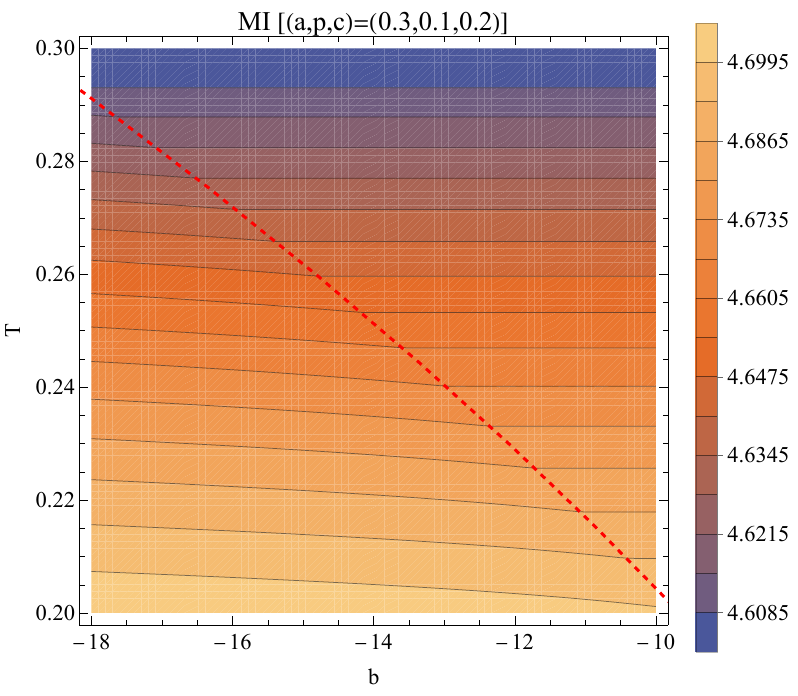}
\caption{The behavior of the renormalized HEE and MI versus temperature $T$ and coupling constant $b$ when the phase transition occurs. The red dashed line represents the critical point. Left plot: the behavior of the renormalized HEE $S_E$ when we set the width $w=1$. After subtracting the divergent term, $S_E$ can take negative values. Right plot: the behavior of MI when we set the configurations $(a,p,c)=(0.3,0.1,0.2)$.}
\label{fig:heemi}
\end{figure}

It needs to be noted that entanglement entropy is positive by definition. However, the asymptotic AdS boundary introduces a divergence in the HEE. In this paper, we subtract this divergent term, allowing the renormalized HEE to take negative values. The left plot of Fig.~\ref{fig:heemi} displays the renormalized HEE behavior across the phase transition. The renormalized HEE effectively signals the transition: it decreases sharply as the coupling constant $b$ diminishes and the scalarized solution emerges, with a similar sharp drop as temperature $T$ falls through the critical point. The rate of change accelerates markedly in the scalarized phase. However, for large strip widths, the renormalized HEE becomes dominated by thermal entropy \cite{Ling:2015dma, Ling:2016wyr}---consistent with the parallel behavior of the renormalized HEE and thermal entropy as $T$ decreases in EMS theory. This thermal contamination limits the utility of the renormalized HEE for probing genuine mixed-state entanglement, motivating the use of alternative measures.

The right plot of Fig.~\ref{fig:heemi} shows MI behavior during the transition. In stark contrast to the renormalized HEE, MI increases as $b$ decreases and rises sharply at the critical point; it also grows with decreasing $T$. This opposite trend suggests that MI better captures the quantum correlations enhanced during the transition. Nevertheless, in certain configurations MI remains influenced by thermal contributions \cite{Huang:2019zph}, warranting examination of additional mixed-state measures.

Having examined the static entanglement measures HEE and MI, we now turn to EWCS and the dynamical measure, butterfly velocity $v_B$.

\subsection{The entanglement wedge cross section and the butterfly velocity}
\begin{figure}
  \centering
  \includegraphics[width=0.45\textwidth]{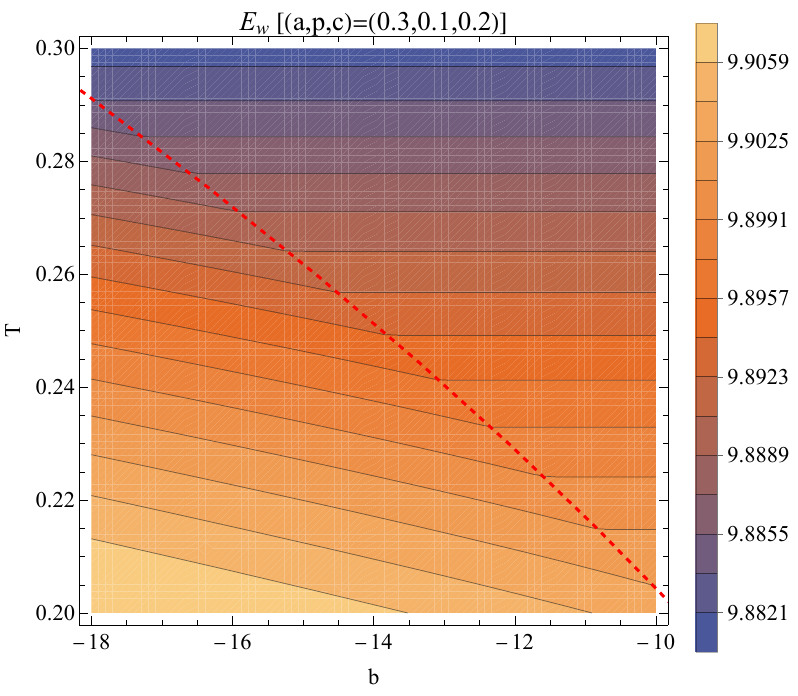}\quad
  \includegraphics[width=0.45\textwidth]{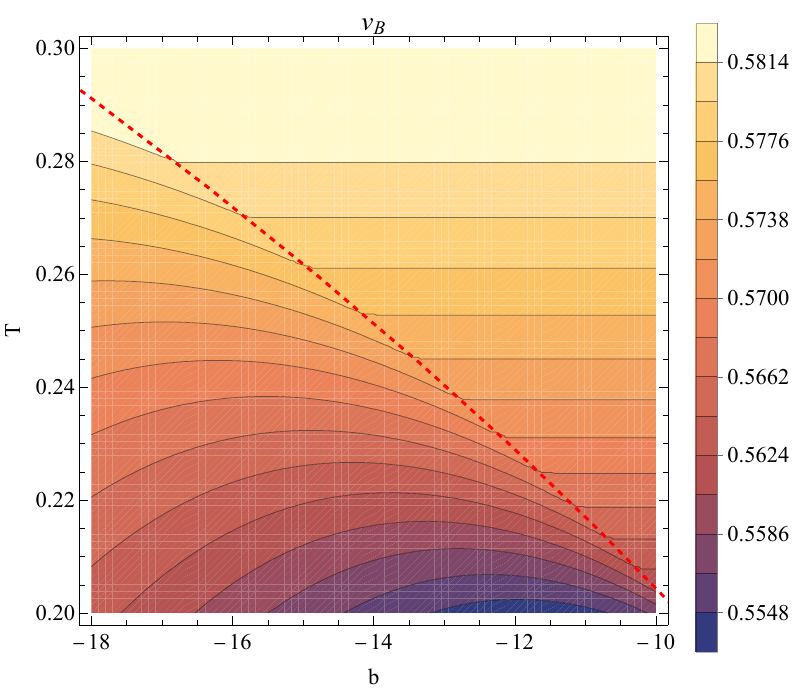}
  \caption{The red dashed line represents the critical point of the phase transition. Left plot: the behavior of EWCS $E_w$ versus different coupling constant $b$ and the temperature $T$. Right plot: the behavior of butterfly velocity $v_B$ versus different coupling constant $b$ and the temperature $T$.}
  \label{fig:ewcsvb}
\end{figure}

The left plot of Fig.~\ref{fig:ewcsvb} presents the EWCS behavior. Like MI, EWCS increases as $b$ decreases and as $T$ drops, exhibiting behavior opposite to HEE. This confirms that both MI and EWCS, as mixed-state measures, capture aspects of entanglement that HEE misses.

The right plot reveals strikingly different behavior for the dynamical measure $v_B$. Unlike the monotonic trends of static measures, $v_B$ displays non-monotonic behavior with respect to $b$: it first decreases then increases as $b$ diminishes. In contrast, $v_B$ decreases monotonically with temperature. This non-monotonicity arises from competing contributions in Eq.~\eqref{eqt:vb}. Specifically,
\begin{equation}
v_B \propto \frac{1}{2V(z)-V'(z)}\bigg|_{z=1}.
\label{eq:vban}
\end{equation} 
The two terms in Eq.~\eqref{eq:vban} encode distinct physics. The thermal entropy density is
\begin{equation}
s=\frac{2\pi A}{\kappa^2}=\frac{2\pi V(z)}{\kappa^2}\hat{V}\bigg|_{z=1},
\label{eq:entropy}
\end{equation}
where $ A $ represents the area of the horizon and $ \hat{V} = \int dx dy $ denotes the area of the corresponding region of the dual system. Consequently, the first term $ V(z) $ in Eq.~\eqref{eq:vban} is associated with the thermal entropy of the system. Moreover, when considering the symmetric EWCS and combining Eq.~\eqref{eq:ewcs} with the ansatz Eq.~\eqref{eq:ansatz}, the expression for the $ E_w $ integral can be expanded as follows,
\begin{equation}
E_w=\int_\Sigma \left( \frac{g_{zz}(z)V'(z)+V(z)g_{zz}'(z)}{4 \sqrt{V(z) g_{zz}(z)}}(z-1)^2+\mathcal{O}(z-1)^3\right)dz.
\label{eq:analyew}
\end{equation}
Therefore, we find that EWCS, as a novel measure of mixed-state entanglement, is given by $ E_w \propto V'(z) $, which corresponds to the second term in Eq.~\eqref{eq:vban}.

\begin{figure}
\centering
\includegraphics[width=0.45\textwidth]{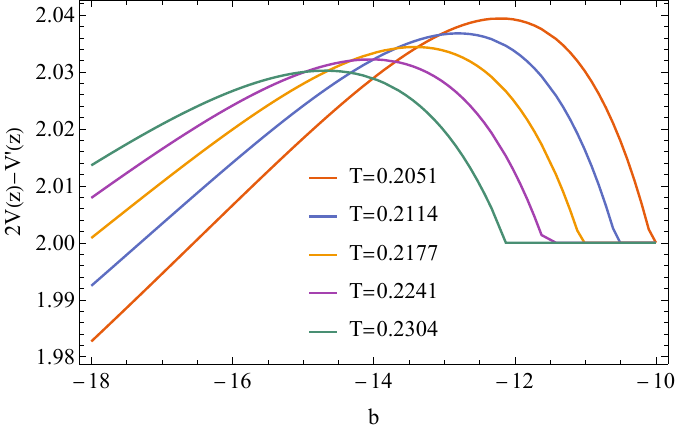}\quad
\includegraphics[width=0.45\textwidth]{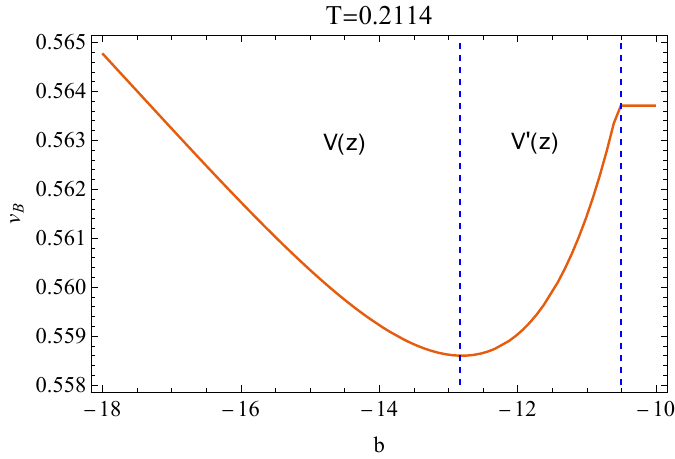}
\caption{Left plot: the function $(2V(z) - V'(z))|_{z=1}$ versus the coupling constant $b$ at different temperatures $T$. Right plot: the butterfly velocity $v_B$ versus $b$ at $T=0.2114$. The blue dashed line represents the peak point of $v_B$ and the critical point, respectively. As the coupling constant $b$ decreases, the region of $v_B$ is predominantly influenced by $V'(z)$ and $V(z)$, respectively.
}
\label{fig:vbfun}
\end{figure}

The left plot of Fig.~\ref{fig:vbfun} shows that $2V(z) - V'(z)$ exhibits non-monotonic behavior due to competition between its two terms. As $b$ decreases, $V'(z)$ initially dominates, causing the function to increase; beyond the peak, $2V(z)$ takes over and the function decreases.

The right plot of Fig.~\ref{fig:vbfun} illustrates how this competition shapes $v_B$. The decreasing portion of $v_B$ is dominated by $V'(z)$, whose near-horizon behavior parallels that of the EWCS integrand (cf. Eq.~\eqref{eq:analyew}), suggesting a potential correlation with mixed-state entanglement. It should be noted, however, that this correlation is inferred from the analytic structure and provides only indirect evidence for the relationship. In contrast, the ascending portion is dominated by $V(z)$, which is directly related to the thermal entropy density via Eq.~\eqref{eq:entropy}.

In summary, the non-monotonic behavior of $v_B$ arises from the competition between two contributions in Eq.~\eqref{eq:vban}: the $V'(z)$ term dominates initially as $b$ decreases, yielding a decrease in $v_B$, while the $V(z)$ term takes over near the critical point, causing $v_B$ to rise. The correspondence of $V(z)$ and $V'(z)$, suggested by Eqs.~\eqref{eq:entropy} and~\eqref{eq:analyew} respectively, provides a phenomenological interpretation of this competition. This intrinsic competition distinguishes the dynamical measure $v_B$ from static entanglement probes and should manifest in other phase transitions as well.

The above analysis reveals distinct behaviors of various entanglement measures during the phase transition. To gain deeper insights into their universal properties, we now investigate their scaling behavior near the critical point.

\subsection{The scaling behavior of the quantum information}\label{subsec:E} 
\begin{figure}
  \centering
  \includegraphics[width=0.45\textwidth]{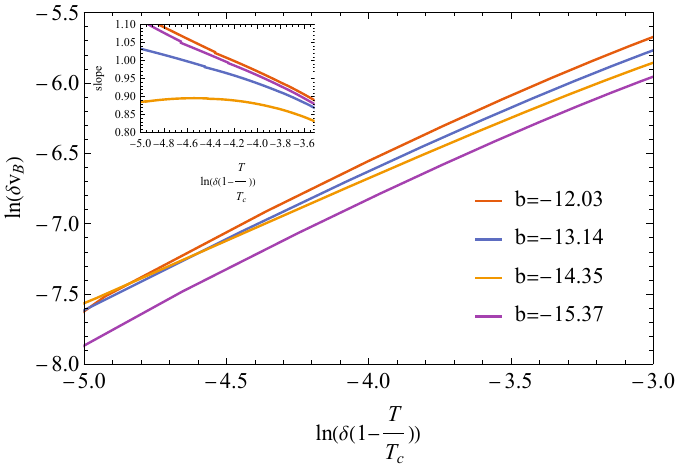}\quad
  \includegraphics[width=0.45\textwidth]{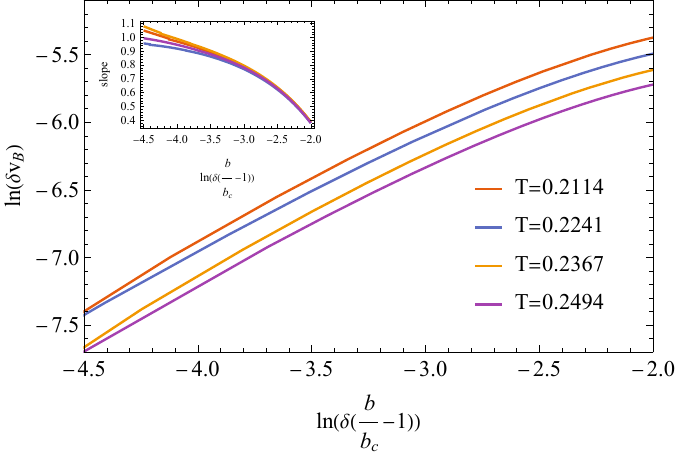}
  \caption{The scaling behavior of butterfly velocity $v_B$. The inset plot is the slope of the scaling behavior, which represents the critical exponent. Left plot: the scaling behavior of $v_B$ with the change of temperature $T$. Right plot: the scaling behavior of $v_B$ with the change of coupling constant $b$, where $b_c$ represents the critical coupling constant.}
  \label{fig:scavb}
\end{figure}

\begin{figure}
  \centering
  \includegraphics[width=0.45\textwidth]{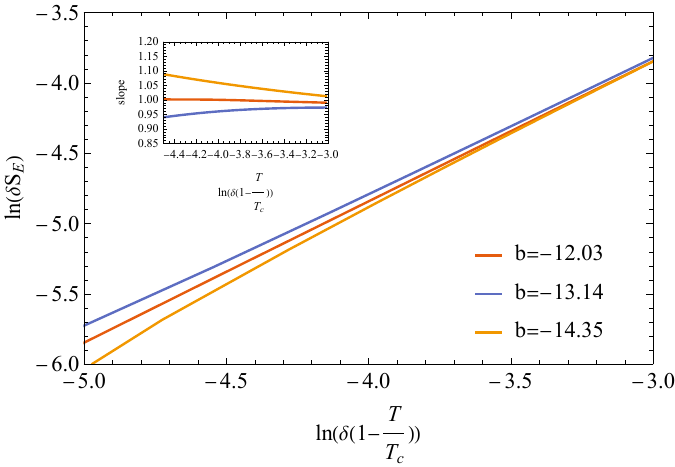}\quad
  \includegraphics[width=0.45\textwidth]{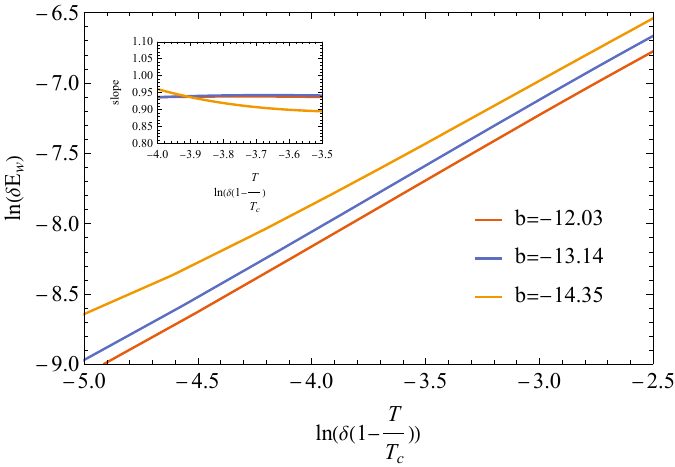}
  \caption{The scaling behavior of HEE $S_E$ and EWCS $E_w$. The inset plot is the slope of the scaling behavior, which represents the critical exponent. Left plot: the scaling behavior of $S_E$ with the change of temperature. Right plot: the scaling behavior of $E_w$ with the change of temperature.}
  \label{fig:scaheeew}
\end{figure}

The critical point in EMS theory marks a second-order phase transition from the normal to the scalarized state. Universality dictates that diverse physical systems share similar scaling behaviors near criticality, making the extraction of critical exponents a valuable diagnostic. We define the difference between scalarized and normal phases:
\begin{equation}
\delta v_B=v_B^\text{scalar}-v_B^{\text{normal}},\quad S_E=S_E^\text{scalar}-S_E^{\text{normal}},\quad \delta E_w=E_w^\text{scalar}-E_w^{\text{normal}}.
\end{equation}
Therefore, the critical behavior for the butterfly velocity, HEE, and EWCS can be read as
\begin{equation}
  \delta v_B\sim\left (1-\frac{T}{T_c}\right )^{\alpha_{v_B}},\quad \delta S_E\sim\left (1-\frac{T}{T_c}\right )^{\alpha_{S_E}},\quad \delta E_w\sim\left (1-\frac{T}{T_c}\right )^{\alpha_{E_w}},
\end{equation}
The critical exponents $\alpha_{v_B}$, $\alpha_{S_E}$, and $\alpha_{E_w}$ characterize the scaling of $v_B$, HEE, and EWCS, respectively. Figure~\ref{fig:scavb} displays the scaling of $v_B$ with both $T$ and $b$; the slope yields the critical exponent. We find 
\begin{equation}
\alpha^T_{v_B}\approx \alpha^b_{v_B}\approx 1.
\end{equation}
In Fig. \ref{fig:scaheeew}, we illustrate the scaling behavior of both the HEE and the EWCS. It is important to note that both the dynamic quantum information and the static quantum information, exhibit similar critical exponents. Our numerical results indicate that,

\begin{equation}
  \alpha_{v_B}\approx\alpha_{S_E}\approx\alpha_{E_w}\approx 1.
\end{equation}

\begin{figure}
  \centering
  \includegraphics[width=0.45\textwidth]{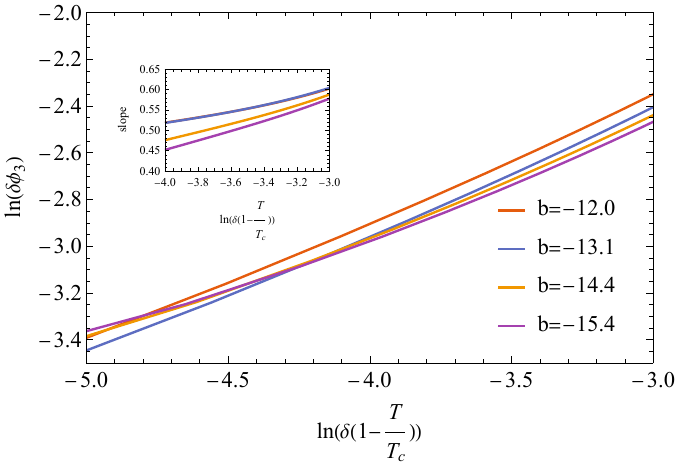}\quad
  \includegraphics[width=0.45\textwidth]{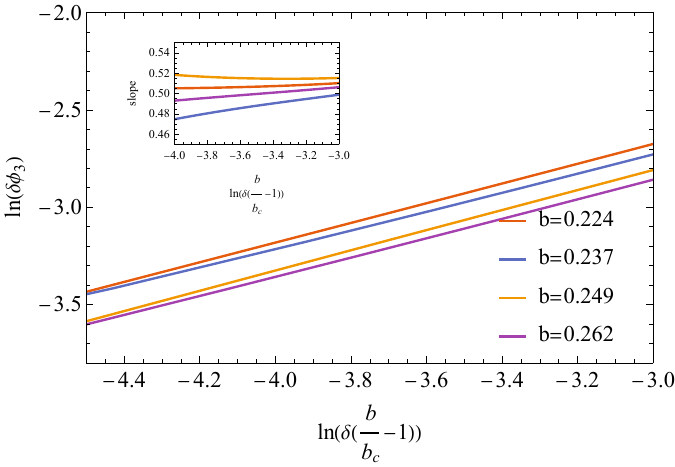}
  \caption{The scaling behavior of $\phi_3$. The inset plot represents the slope of the scaling behavior, which can represent the critical exponent. Left plot: the scaling behavior of $\phi_3$ as a function of temperature. Right plot: the scaling behavior of $\phi_3$ with respect to the coupling constant $b$.}
  \label{fig:scaphi}
\end{figure}

To better investigate why all holographic quantum information exhibit the same critical exponent, we study the scaling behavior of the scalar field $\phi$. The expansion of $\phi$ near the AdS boundary is given by \cite{Zhang:2021etr},
\begin{equation}
\phi=z^3\phi_3+\mathcal{O}(z^4).
\end{equation}
We show the scaling behavior of the $\phi_3$ in Fig. 
\ref{fig:scaphi}. Our numerical result shows that the scaling behavior of the $\phi_3$ with temperature and coupling constant is 
\begin{equation}
\delta(\phi_3)\sim (1-\frac{T}{T_c})^{\alpha^T_\phi}\sim (\frac{b}{b_c}-1)^{\alpha^b_\phi},
\end{equation} 
where the critical exponent $\alpha^T_\phi=\alpha^b_{\phi}=1/2$. It is important to note that the critical exponent of other holographic quantum information is always twice that of $\phi_3$,
\begin{equation}
\alpha_{v_B}=\alpha_{S_E}=\alpha_{E_w}=2\alpha^T_\phi=2\alpha^b_\phi.
\end{equation}
We can consider $\delta \phi_3$ as the perturbation when the phase transition occurs. Near the critical point, we can expand the scalar field and metric function as follows \cite{Zeng:2010zn,Pan:2012jf},
\begin{equation}
  \begin{aligned}
\phi&=\epsilon\phi_{1}+\epsilon^3\phi_{2}+\epsilon^5\phi_{3}+\cdots,\\
U&=1+\epsilon^2 U_2+\epsilon^4 U_3+\cdots,\\
V&=1+\epsilon^2 V_2+\epsilon^4 V_3+\cdots.
\end{aligned}
\end{equation}
We find that the perturbation of the metric function is twice than the scalar field,
\begin{equation}
\delta g_{\mu\nu}\sim (\delta \phi)^2.
\end{equation}
Therefore, we can conclude that the critical exponents of geometry-related quantities are always twice that of the scalar field.

Beyond the individual behavior of each entanglement measure, we are particularly interested in the relative growth rates of MI and EWCS, which may reveal universal features of thermodynamic phase transitions.

\section{The growth rate of the holographic quantum information}
\label{sec:ineq}

Inequalities between different holographic quantum information have been widely studied, such as the inequality $E_w(\rho_{AC}) \geq \frac{1}{2}I(A, C)$ \cite{Bao:2017nhh}. These inequalities can deepen our understanding of the properties of mixed-state entanglement measures. In our previous study, we have found an inequality between the growth rate of the EWCS and the MI in the holographic p-wave superconductor \cite{Yang:2023wuw}. We found that the growth rate of MI near the critical point is always greater than that of EWCS. We propose that this inequality is linked to the definition of holographic quantum information. However, the EMS theory differs from the holographic superconductor theory, and the coupling constant $b$ also influences the phase transition. Therefore, it is necessary to investigate whether the inequality still holds in this theory, which could help broaden the scope of where such an inequality occurs.

When the temperature or coupling constant crosses the critical point, the system occurs the phase transition from the normal state to the scalarized state. To better investigate the growth rate of the holographic quantum information, we define the relative values of the MI and the EWCS as follows:
\begin{equation}
\tilde{I}=\frac{I_\text{scalar}}{I_\text{norm}}, \quad \tilde{E_w}=\frac{E_{w,\text{scalar}}}{E_{w,\text{norm}}}.
\end{equation}
With this definition, we can fix $\tilde{E_w}$ and $\tilde{I}$ to $1$ at the critical point. As the temperature or the coupling constant changes, the relative values of MI and EWCS begin to change rapidly. Near the critical point, the behavior of these relationships can be described as follows:
\begin{equation}
\delta(Q)\sim A(Q)(1-\frac{T}{T_c})^\alpha,
\end{equation}
where $Q$ is any physical quantity that exhibits critical behaviors. Therefore, EWCS and MI can be represent as,

\begin{equation}
  \tilde{E_w}=1+A(\tilde{E_w})\left(1-\frac{T}{T_c}\right)^\alpha,\quad  \tilde{I}=1+A(\tilde{I})\left(1-\frac{T}{T_c}\right)^\alpha.
\end{equation}

\begin{figure}
  \centering
  \includegraphics[width=0.55\textwidth]{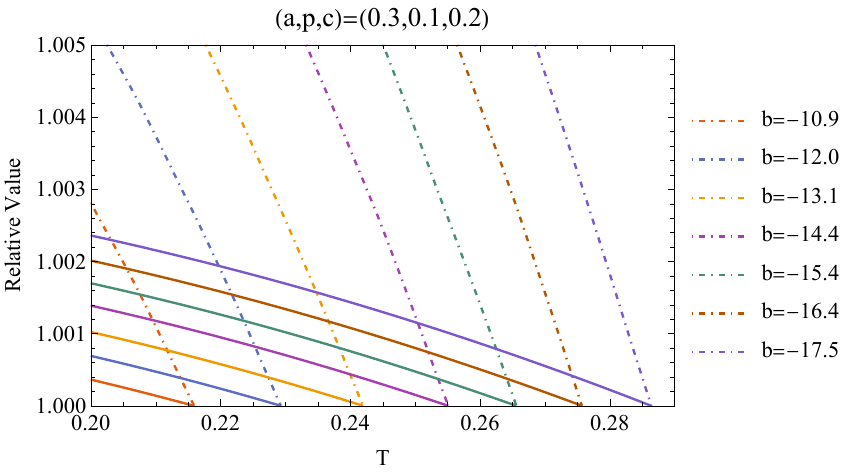}\quad
  \includegraphics[width=0.35\textwidth]{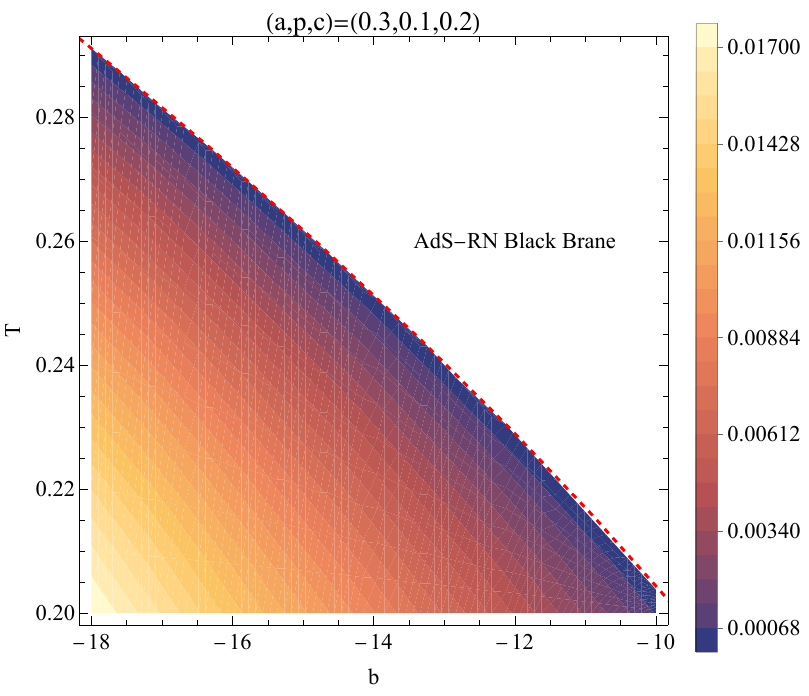}
  \caption{The growth rate of EWCS and MI with the same configuration. Left plot: the relative values of EWCS and MI with varying coupling constants $ b $, at the phase transition point. The solid line represents the relative value of EWCS $\tilde{E_w}$. The dashed line represents the relative value of MI $\tilde{I}$. Right plot: the value of $ A(\tilde{I}) - A(\tilde{E_w}) $. The red dashed line marks the critical point of the phase transition.}
  \label{fig:ineq}
  \end{figure}
  It is easy to observe that $ A(Q) $ measures the increasing behavior of holographic quantum information, which allows us to refer to $ A $ as the growth rate. In the left plot of Fig. \ref{fig:ineq}, we present the relative values of EWCS and MI at the phase transition. We find that the relative value of MI increases more quickly than that of EWCS, indicating that the growth rate of MI is higher than EWCS,
\begin{equation}
  A(\tilde{I})\geq A(\tilde{E_w}).
\end{equation}
To better investigate this inequality in holographic quantum information, we show the value of $ A(\tilde{I}) - A(\tilde{E_w}) $ in the right plot of Fig. \ref{fig:ineq}. We find that whenever the phase transition occurs, either with respect to temperature $ T $ or the coupling constant $ b $, the inequality is consistently validated. Additionally, as one moves away from the critical point, the inequality becomes more significant. To address the dependence on subsystem configurations, we computed the critical exponents and the growth rate inequality for various configurations. The results are summarized in Table~\ref{tab:ineq}. For MI and EWCS, we considered several different $(a,p,c)$ configurations, and the critical exponents remain $\alpha \approx 1$ within numerical precision. Furthermore, the inequality $A(\tilde{I}) > A(\tilde{E}_w)$ holds universally across all configurations examined, with the growth rate difference $A(\tilde{I}) - A(\tilde{E}_w)$ remaining positive. These results confirm that our findings are independent of the subsystem configuration and reflect the universal features of the model. Therefore, we propose that this inequality is closely related to the thermodynamic phase transition. Based on our previous investigation of the holographic p-wave superconductor \cite{Yang:2023wuw}, we demonstrate that this inequality generally holds within this class of theories. This inequality appears in the system whenever a thermodynamic phase transition occurs. We suggest that this inequality is associated with the properties of holographic quantum information. MI measures the total correlation of the system, which comprises both quantum and classical correlations. On the other hand, EWCS is considered as the holographic dual of several different mixed-state entanglement measures, which only capture part of the correlation within the system. Therefore, when the phase transition occurs, despite having the same configuration, MI can capture more information than EWCS, making it more significant in diagnosing the phase transition.

This inequality deepens our understanding of the relationship among holographic entanglement measures and provides a window into the universal features of thermodynamic phase transitions.

\begin{table}[htbp]
\centering
\caption{The critical exponents $\alpha_I$, $\alpha_{E_w}$ and the growth rates $A(\tilde{I})$, $A(\tilde{E}_w)$ for different subsystem configurations at $b=-13.1$.}
\label{tab:ineq}
\begin{tabular}{| @{\hskip 15pt}c@{\hskip 15pt} | @{\hskip 15pt}c@{\hskip 15pt}| @{\hskip 15pt}c@{\hskip 15pt} | @{\hskip 15pt}c@{\hskip 15pt} | @{\hskip 15pt}c@{\hskip 15pt} |}
\hline
Configuration & $\alpha_{I}$ & $\alpha_{E_w}$ &  $A(\tilde{I})$ & $A(\tilde{E}_w)$ \\
\hline
$(a,p,c)=(0.5,0.1,0.4)$ & 0.99 & 1.04 & 0.05685 & 0.01504 \\ \hline
$(a,p,c)=(0.8,0.1,0.7)$ & 1.01 & 1.06 & 0.09694 & 0.02509 \\ \hline
$(a,p,c)=(1.0,0.1,0.9)$ & 1.02 & 1.04 & 0.12098 & 0.02736 \\ \hline
$(a,p,c)=(0.8,0.1,0.3)$ & 1.00 & 1.04 & 0.06841 & 0.01831 \\ \hline
$(a,p,c)=(1.0,0.1,0.3)$ & 1.00 & 1.04 & 0.07753 & 0.02105 \\ \hline
$(a,p,c)=(1.2,0.1,0.3)$ & 1.01 & 1.04 & 0.08469 & 0.02301 \\ \hline
$(a,p,c)=(1.5,0.1,0.3)$ & 1.01 & 1.04 & 0.09164 & 0.02494 \\ \hline
$(a,p,c)=(0.5,0.25,0.4)$ & 0.99 & 1.05 & 0.42733 & 0.03786 \\ \hline
$(a,p,c)=(0.7,0.3,0.6)$ & 1.00 & 1.07 & 2.26141 & 0.08395 \\ \hline
$(a,p,c)=(0.9,0.32,0.8)$ & 1.01 & 1.05 & 1.81762 & 0.09575 \\ \hline
$(a,p,c)=(0.2,0.1,0.15)$ & 0.98 & 0.98 & 0.19432 & 0.00368 \\ \hline
$(a,p,c)=(0.15,0.06,0.1)$ & 0.95  & 0.97  & 0.01301 & 0.00012 \\ 
\hline
\end{tabular}
\end{table}

\section{Discussion}\label{sec:4}

We have systematically investigated holographic entanglement measures in EMS theory, encompassing both static probes (HEE, MI, EWCS) and the dynamical butterfly velocity $v_B$. All measures effectively diagnose the phase transition, yet they exhibit qualitatively distinct behaviors. While HEE decreases with both decreasing temperature and coupling constant, the mixed-state measures MI and EWCS increase---highlighting their complementary sensitivity to quantum correlations. The butterfly velocity $v_B$ behaves differently still: it varies monotonically with temperature but non-monotonically with the coupling constant. This non-monotonicity reflects a competition between contributions of the space metric $V(z)$, offering insight into how dynamical measures probe phase transitions. We note that the separation of $v_B$ into entanglement-dominated and thermal-dominated regimes, while physically intuitive, is established through the near-horizon behavior of the metric function $V(z)$ and its potential parallels with thermal entropy and the mixed-state entanglement. However, a rigorous first-principles decomposition of $v_B$ into entanglement and thermal contributions is still lacking, and the interpretation should be viewed as a phenomenologically motivated diagnostic. Developing such a decomposition would be an important direction for future work.

We also examined the scaling behavior near criticality. All entanglement measures share a universal critical exponent $\alpha = 1$, precisely twice that of the scalar order parameter ($\alpha_\phi = 1/2$). This doubling arises because the entanglement measures depend quadratically on the scalar field. Furthermore, we identified an inequality between the growth rates of MI and EWCS: MI consistently grows faster than EWCS during the transition. Since MI captures total correlations while EWCS measures only a subset, this hierarchy appears intrinsic to holographic systems and should persist across thermodynamic phase transitions.

Looking ahead, extending this analysis to other phase transitions---topological, quantum, and in alternative gravity theories \cite{Zhang:2020kxz,Liu:2020evp,Liu:2020lwc,Donos:2013eha, Landsteiner:2015pdh, Ling:2016dck, Baggioli:2018afg, Baggioli:2020cld,Ling:2015exa,Ling:2015epa,Ling:2014bda}---presents a natural next step. In particular, it remains an open question whether the MI--EWCS inequality holds for quantum phase transitions, whose underlying mechanisms differ fundamentally from thermodynamic ones.

\section*{Acknowledgments}
Peng Liu would like to thank Yun-Ha Zha, Yi-Er Liu and Bai Liu for their kind encouragement during this work. Zhe Yang appreciates Feng-Ying Deng's support and warm words of encouragement during this work. We are also very grateful to Chong-Ye Chen for their helpful discussion and suggestions. This work is supported by the Natural Science Foundation of China under Grant Nos. 12475054, 12375055 and the Guangdong Basic and Applied Basic Research Foundation No. 2025A1515012063. Zhe Yang is supported by the Jiangsu Postgraduate Research and Practice Innovation Program under Grant No. KYCX25\_3922.

\nocite{*}
	
\end{document}